\documentclass[prd,reprint,superscriptaddress,preprintnumbers,eqsecnum,showpacs,nofootinbib,nobibnotes]{revtex4-1}
\usepackage{amsfonts,amsmath,amssymb,bm,natbib}
\usepackage{graphicx}
\usepackage{slashed} 
\usepackage{subfigure}
\usepackage{color}

\newcommand{\be}{\begin{equation}}
\newcommand{\bea}{\begin{eqnarray}}
\newcommand{\ee}{\end{equation}}
\newcommand{\eea}{\end{eqnarray}}

\def\s#1{{\scriptscriptstyle #1}}

\def\1eq#1{Eq.(\ref{#1})}

\def\2eqs#1#2{Eqs.~(\ref{#1}) and~(\ref{#2})}
\def\3eqs#1#2#3{Eqs.~(\ref{#1}),~(\ref{#2}) and~(\ref{#3})}

\newcommand{\STI}{\textnormal{\tiny \textsc{STI}}}
\newcommand{\Tr}{\textnormal{\tiny \textsc{T}}}
\newcommand{\FBC}{\rm{\s{FBC}}}
\newcommand{\BC}{\rm{\s{BC}}}

\begin{document}

\title{Quark mass generation with Schwinger-Dyson equations}
\author{A.~C. Aguilar}
\email{Invited talk given by A. C. Aguilar at the 
``XIV International Workshop on Hadron Physics'', 18-23 March 2018, Florian\'opolis, Brazil.}
\affiliation{University of Campinas - UNICAMP,  \\
Institute of Physics ``Gleb Wataghin'',
13083-859 \\ Campinas, SP, Brazil}

\author{M.~N. Ferreira}
\affiliation{University of Campinas - UNICAMP,  \\
Institute of Physics ``Gleb Wataghin'',
13083-859 \\ Campinas, SP, Brazil}

\begin{abstract}

In this talk, we review  some of the current efforts to understand the phenomenon of chiral symmetry breaking and the generation of a dynamical quark mass.  To do that, we will use the standard framework of the Schwinger-Dyson equations. The key ingredient in this analysis is the quark-gluon vertex, whose 
non-transverse part may  be determined exactly  from   the  nonlinear   Slavnov-Taylor  identity   that  it satisfies. The resulting expressions for the form factors of this vertex  involve  not only  the quark
propagator, but also the ghost dressing function and the quark-ghost
kernel.  Solving the coupled system of integral equations formed by the quark propagator and the four form factors of the scattering kernel, we carry out  a detailed  study  of the  impact  of the quark-gluon  vertex  on the  gap equation and the quark masses  generated from it, putting particular
 emphasis on the contributions directly related with the ghost sector
of  the   theory,  and   especially  the  quark-ghost   kernel.  Particular attention is dedicated on the way that the correct renormalization group behavior of the dynamical quark mass is recovered, and in the extraction of the phenomenological parameters such as the pion decay constant. 

\end{abstract}


\maketitle

\section{\label{sec:intro} Introduction}

The dynamical chiral symmetry breaking and the subsequent mass generation for the quarks are eminently nonperturbative phenomena, and they have been the central focus of a series of studies~\cite{Roberts:1994dr,Maris:1999nt,Fischer:2003rp,Maris:2003vk,Aguilar:2005sb,Bowman:2005vx,Aguilar:2010cn,Cloet:2013jya,Mitter:2014wpa,Binosi:2016wcx}. One of the main nonperturbative tools to investigate these phenomena is the Schwinger-Dyson equation (SDE) for the quark propagator, often called the ``quark gap equation''.

In the framework of SDEs, the self consistent truncation of the infinite system of coupled integral equations poses the major difficulty. For the quark gap equation, the challenge mainly consists of constructing an Ansatz for the quark-gluon vertex, $\Gamma_{\mu}(q,p_2,-p_1)$, a complicated three point function composed by twelve linearly independent tensor structures~\cite{Ball:1980ay,Kizilersu:1995iz,Davydychev:2000rt}. More specifically, each one of the twelve tensorial structures are accompanied by its respective form factor. The latter are functions of three-variables,
chosen to be the moduli of two of the incoming momenta, $p_1$ and $p_2$, and the angle $\theta$ between them.
 
 Given that the quark propagator is known to be rather sensitive to the details of the quark-gluon vertex entering in the kernel of the gap equation, it is pressing to determine  the nonperturbative behavior of the aforementioned form factors. 

One strategy to determine part of the twelve form factors of the quark-gluon vertex nonperturbatively was put forth in~\cite{Aguilar:2010cn,Aguilar:2016lbe}. There, using the guiding principles of the ``gauge technique'', 
it was shown that the  Slavnov-Taylor identity (STI) that 
$\Gamma_{\mu}$ satisfies, relates the behavior of its form factors  to other three quantities: (i) the quark propagator $S(p)$, (ii) the ghost dressing function $F(q)$, and (iii) the quark-ghost scattering kernel $H(q,k,-p)$. More specifically, out of the twelve form factors, the STI constrains the behavior of four of them, while the other eight, being transverse to the gluon momentum $q$, satisfy the STI trivially and hence are left undetermined from the identity. 

In this talk, we will discuss the construction of a 
set of  coupled integral equations governing the dynamics of the quark propagator, $S(p)$, and the scattering kernel, $H(q,k,-p)$ in the Landau gauge.  Then, using the STI, the behavior of the four non-transverse form factors of $\Gamma_{\mu}$ will be determined. Finally, we will present 
 our numerical results and discuss the  impact of the these form factors on the  dynamical quark mass generation~\cite{Aguilar:2018epe}.

\section{\label{sec:background} The system of coupled equations}

\begin{figure}[!h]
\includegraphics[width=1.\columnwidth]{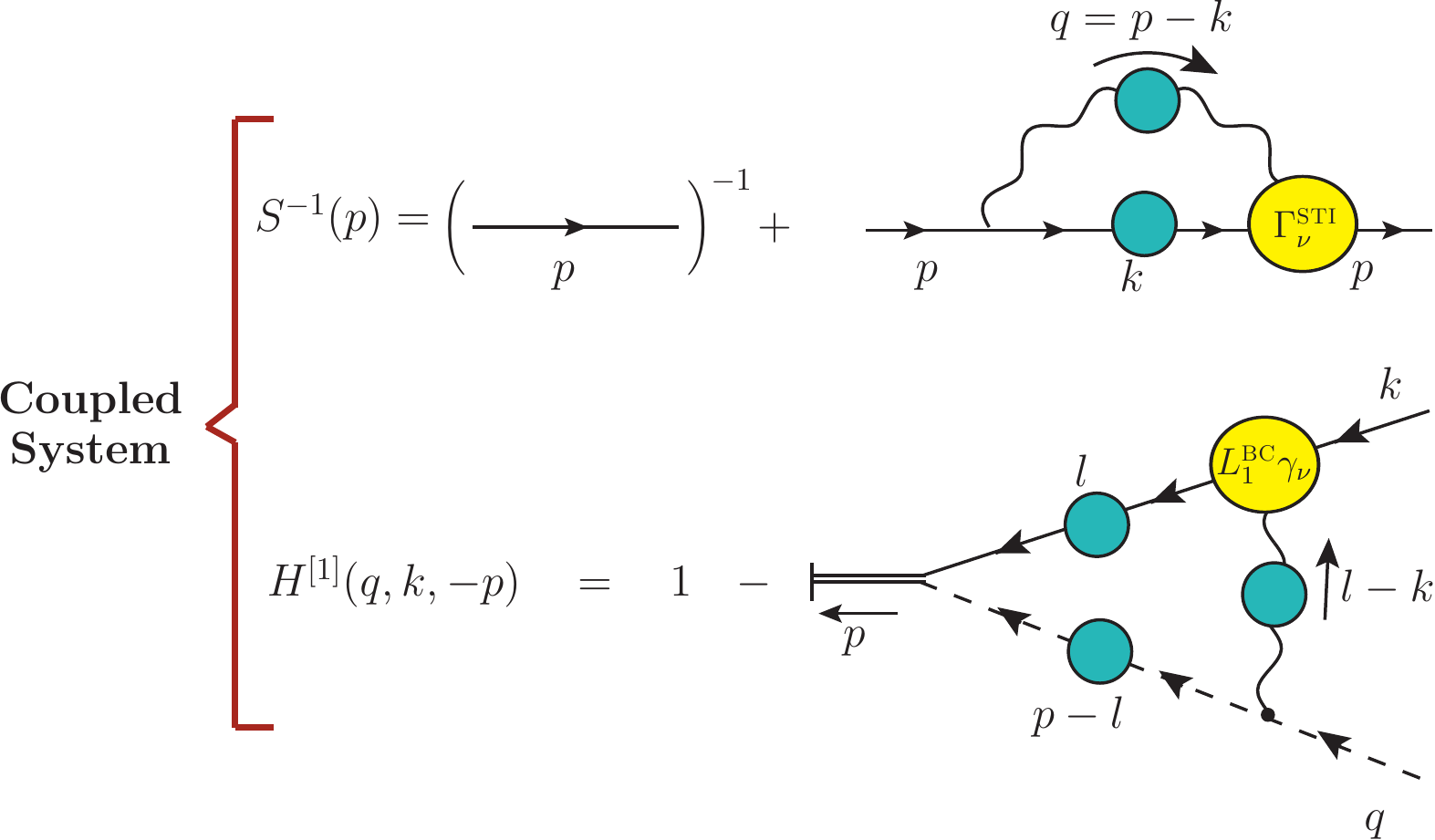}
\vspace{-0.5cm}
\caption{\label{fig:system} Diagrammatic representation of the SDEs for the quark propagator, $S(p)$ (top), and 
the one-loop dressed truncation of the quark-ghost scattering kernel, \mbox{$H^{[1]}(q,k,-p)$} (bottom). The two equations are coupled via the quark-gluon vertex, $\Gamma^{\STI}_{\mu}$, related to $S(p)$ and \mbox{$H^{[1]}(q,k,-p)$} through a STI.}
\end{figure}

The coupled system of SDEs for $S(p)$ and $H(q,k,-p)$ which will be the 
central focus of the present work is shown diagrammatically in Fig.~\ref{fig:system}. 
The full quark propagator  can be written as
\be
S^{-1}(p) =  A(p)\slashed{p} - B(p) \mathbb{I} 
= A(p)[\slashed{p}-{\mathcal{M}}(p) \mathbb{I}] \,,
\label{qAB}
\ee
where $A^{-1}(p)$ is the quark wave function, and the  pole of the
propagator, \mbox{$\mathcal{M}(p) = B(p)/A(p)$}, defines the dynamical quark mass.

The renormalized version of the  quark gap equation appearing on the top of Fig.~\ref{fig:system} may be written as (in the chiral limit)
\begin{equation}
S^{-1}(p)= Z_{\rm{\s F}}  \slashed{p}  -Z_{1}C_{\rm{\s F}}g^2\!\!\int_k\,
\!\!\gamma_{\mu}S(k)\Gamma_{\nu}\Delta^{\mu\nu}(q) \,,
\label{gap}
\end{equation}%
where $C_{\rm {\s F}}$ denotes the Casimir eigenvalue for the fundamental representation, while $Z_1$ and $Z_{\rm{\s F}}$ are the quark-gluon vertex and the quark wave function renormalization constants. 
Our analysis, we will be carried out in the Landau gauge. This is the most common choice, because the entire gluon propagator is transverse, both its self-energy and its free part, whereas for any other value of the gauge-fixing parameter the free part is not transverse. Other non covariant gauges can be also used to study
chiral symmetry breaking, such as Coulomb gauge~\cite{Pak:2011wu}. 

 Therefore, in Landau gauge, 
the gluon propagator reads
\begin{align}
i\Delta_{\mu\nu}(q)=-iP_{\mu\nu}(q)\Delta(q); \;\; P_{\mu\nu}(q)=g_{\mu\nu}-\frac{q_\mu q_\nu}{q^2} \,.
\label{propagators}
\end{align}
In addition, we have introduced the compact notation $\int_{k}\equiv {\mu^{\epsilon}}/{(2\pi)^{d}}\!\int\!\mathrm{d}^d k$, where $\mu$ is the 't Hooft mass, and  $d=4-\epsilon$ is the space-time dimension.

The quark-gluon vertex $\Gamma_\mu$, appearing in Eqs.~\eqref{gap}, is commonly split in the following way
\be
\Gamma_{\mu}(q,p_2,-p_1)=\Gamma^{\STI}_{\mu}(q,p_2,-p_1)+\Gamma^{\Tr}_{\mu}(q,p_2,-p_1)\,,
\label{split}
\ee
where the transverse part, $\Gamma^{\Tr}_{\mu}$, is automatically conserved, \emph{i.e.}
\be
q^{\mu}\Gamma^{\Tr}_{\mu}(q,p_2,-p_1)=0\,, 
\ee
whereas $\Gamma^{\STI}_{\mu}$ (non-transverse) satisfies the STI given by
\begin{align}
 q^{\mu}\Gamma^{\STI}_{\mu}(q,p_2,-p_1)=& F(q)[S^{-1}(p_1) H -{\overline H}S^{-1}(p_2)] \,,
\label{STI}
\end{align}
where $F(q)$ is the ghost dressing function, defined
in terms of ghost propagator as \mbox{$D(q)=iF(q)/q^2$}. In addition, in the
STI, appears the quark-ghost scattering kernel 
$H(q,p_2,-p_1)$, diagrammatically represented in the bottom panel of Fig.~\ref{fig:system}. Notice that for the sake of
notational compactness, 
we have omitted the functional dependences of both $H(q,p_2,-p_1)$
and its ``conjugate'' \mbox{${\overline {H}}(-q,p_1,-p_2)$}~\cite{Aguilar:2016lbe}. 

The most general Lorentz decomposition for $H$ and ${\overline {H}}$ is 
written as~\cite{Davydychev:2000rt,Aguilar:2010cn,Aguilar:2016lbe}
\begin{align}
H=X_0\mathbb{I}+X_1\slashed{p}_1+X_2\slashed{p}_2+X_3\widetilde{\sigma}_{\mu\nu}p_1^{\mu}p_2^{\nu}\,,\nonumber \\
\overline{H}=\overline{X}_0\mathbb{I}+\overline{X}_2\slashed{p}_1+\overline{X}_1\slashed{p}_2+\overline{X}_3\widetilde{\sigma}_{\mu\nu}p_1^{\mu}p_2^{\nu}\,, 
\label{Hdecomp}
\end{align}
where $\widetilde\sigma_{\mu\nu} \equiv \frac{1}{2}[\gamma_{\mu},\gamma_{\nu}]$, \mbox{$X_i:= X_i(q^2,p_2^2,p_1^2)$}, and \mbox{$\overline{X}_i:= X_i(q^2,p_1^2,p_2^2)$}. At tree level, the only nonzero form factors are $X^{(0)}_0=\overline{X}^{(0)}_0=1$.

Similarly, $\Gamma^{\STI}_{\mu}$ can be decomposed in the Ball-Chiu basis as~\cite{Ball:1980ay}
\begin{align}
\Gamma^{\STI}_{\mu}(q,p_2,\!-p_1\!) &=
  L_1 \gamma_{\mu}
+ L_2 (\slashed{p}_1 - \slashed{p}_2)(p_1-p_2)_{\mu} 
\nonumber\\
&\hspace{-0.25cm} + L_3 (p_1-p_2)_{\mu} 
+ L_4 \tilde\sigma_{\mu\nu}(p_1-p_2)^{\nu} \,,
\label{Li}
\end{align}
where \mbox{$L_i:=L_i(q^2, p_2^2, p_1^2)$}.

Substituting into the STI~\eqref{STI} the Eqs.~\eqref{qAB},~\eqref{Hdecomp}, and~\eqref{Li}, 
it is possible to express \mbox{$L_i:=F(q)\overline{L}_i/2$} as~\cite{Aguilar:2010cn} 
\begin{widetext}
\vspace{-0.3cm}
\bea
\overline{L}_1 &=& 
A(p_1)[X_0 - (p_1^2+p_1\!\cdot\!p_2)X_3] 
+ A(p_2)[{\overline X}_0 -(p_2^2 + p_1\!\cdot\!p_2){\overline X}_3] + B(p_1)(X_2-X_1) + B(p_2)({\overline X}_2-{\overline X}_1)\,;
\nonumber\\
\overline{L}_2 &=& \frac{1}{(p_1^2 - p_2^2)} \left\{
A(p_1)[X_0 + (p_1^2 - p_1\!\cdot\!p_2)X_3] 
- A(p_2)[{\overline X}_0 +(p_2^2-p_1\!\cdot\!p_2){\overline X}_3] - B(p_1)(X_1+X_2) + B(p_2)({\overline X}_1+{\overline X}_2)\right\};
\nonumber\\
\overline{L}_3 &=&  \frac{2}{p_1^2 - p_2^2}
\left\{  
A(p_1) \left( p_1^2 X_1 + p_1\!\cdot\!p_2 X_2 \right)
- A(p_2) \left( p_2^2 {\overline X}_1 +p_1\!\cdot\!p_2 {\overline X}_2\right)
- B(p_1)X_0 + B(p_2){\overline X}_0\right\};
\nonumber\\
\overline{L}_4 &=& 
A(p_1) X_2 - A(p_2) {\overline X}_2 - B(p_1) X_3 + B(p_2){\overline X}_3 \,.
\label{expLi}
\eea
\vspace{-0.7cm}
\end{widetext}

Notice that setting to tree level the various $X_i$, appearing
in Eq.~\eqref{expLi}, we obtain the  so-called ``minimally non-abelianized''
$\Gamma^{\FBC}_{\mu}$~\cite{Fischer:2003rp, Aguilar:2010cn,Aguilar:2016lbe},
whose form factors $L_i^{\FBC} = F(q)L_i^{\BC}$ are given by 
\begin{align}\label{BC}
L_1^{\BC} &= \frac{[A(p_1)+A(p_2)]}{2}\,,\quad L_2^{\BC} = \frac{[A(p_1)- A(p_2)]}{2(p_1^2 - p_2^2)},  \nonumber \\
L_3^{\BC} &= -\frac{[B(p_1)- B(p_2)]}{p_1^2 - p_2^2} \,, \quad  L_4^{\BC} = 0 \,.
\end{align}

In what follows, we will neglect the transverse part of the quark-gluon vertex, \emph{i.e.} we set 
\mbox{$\Gamma^{\Tr} = 0$}, in the gap equation~\eqref{gap}, since it can not be determined from the STI.

To proceed, we substitute into the gap equation~\eqref{gap} the dressed quark-gluon vertex of Eq.~\eqref{Li} using  \mbox{$p_1 = p$}  and \mbox{$p_2 = k$}. After taking the  traces, we arrive in the following expressions for the integral equations satisfied by $A(p)$ and $B(p)$ (in the Euclidean space)~\cite{Aguilar:2010cn} 
\bea\label{gAB}
p^2A(p)  &=& Z_{\rm{\s F}} p^2 + Z_1 4\pi C_{F}\alpha_s\!\int_{k} {\mathcal K}_{\rm{\s A}}(k,p)\Delta(q)F(q)\, , \nonumber\\
 B(p)  &=& Z_1 4\pi C_{F} \alpha_s\! \int_{k} {\mathcal K}_{\rm{\s B}}(k,p)\Delta(q)F(q)\,,
\eea 
where \mbox{$\alpha_s = g^2(\mu)/4\pi$} and the kernels are given by 
\bea
{\mathcal K}_{\rm{\s A}}(k,p)&=&\left\{\frac{3}{2} (k \!\cdot\!p)\overline{L}_{1} -[\overline{L}_{1} -(k^2+p^2) \overline{L}_{2}]h(p,k) \right\}{\mathcal Q}_{\rm{\s A}}(k) \nonumber\\ 
&-& \left\{\frac{3}{2}p \!\cdot\!(k+p)\overline{L}_{4} +(\overline{L}_{3} - \overline{L}_{4})h(p,k)\right\}{\mathcal Q}_{\rm{\s B}}(k)\,,\nonumber\\ 
{\mathcal K}_{\rm{\s B}}(k,p)&=&\left\{\frac{3}{2} k \!\cdot\!(k+p)\overline{L}_{4} - (\overline{L}_{3}+\overline{L}_{4})h(p,k) \right\}{\mathcal Q}_{\rm{\s A}}(k) \nonumber\\ 
 &+& \left\{\frac{3}{2}\overline{L}_{1}  - 2 h(p,k)\overline{L}_{2}\right\}{\mathcal Q}_{\rm{\s B}}(k)\,,
\label{kernelAB}
\eea
with the functions $h(p,k)$ and ${\mathcal Q}_{\rm{\s f}}(k)$ defined as 
\bea
h(p,k) &:=& \frac{\left[k^2p^2-(k\!\cdot\!p)^2\right]}{q^2}\,,\nonumber \\
{\mathcal Q}_{\rm{\s f}}(k) &:=& \frac{f(k)}{[A^2(k)k^2+B^2(k)]}\,,
\label{hf}
\eea
where $f(k)$, appearing in the numerator of  Eq.~\eqref{hf}, can be either $A(k)$ or $B(k)$, depending on the index of  ${\mathcal Q}$.

Next, concerning the renormalization of Eqs.~\eqref{gAB}, we notice that in Landau gauge $S(p)$ and $H(q,k,-p)$ are finite at one loop~\cite{Nachtmann:1981zg}, so that we may set $Z_{\rm{ \s F }} = Z_{\rm{ \s H }} = 1$. In addition, it follows from the STI in Eq.~\eqref{STI} that the renormalization constants are related by $Z_1 = Z_c^{-1} Z_{\rm{ \s F }}Z_{\rm{ \s H }}$, thus in Landau gauge $Z_1 = Z_c^{-1}$.  Applying the 
above constraints in the Eq.~\eqref{gAB}, we obtain
\bea
p^2A(p)  &=& p^2 + Z_{\rm{\s c}}^{-1} 4\pi C_{F}\alpha_s\!\int_{k} {\mathcal K}_{\rm{\s A}}(k,p)\Delta(q)F(q)\, , \nonumber\\
 B(p)  &=& Z_{\rm{\s c}}^{-1} 4\pi C_{F} \alpha_s\! \int_{k} {\mathcal K}_{\rm{\s B}}(k,p)\Delta(q)F(q)\,.
\label{rsenergy1}
\eea%

The presence of $Z_c^{-1}$ multiplying the self-energy in the Eq.~\eqref{rsenergy1} is a final complicating factor to be addressed in the nonperturbative truncation of the gap equation~\cite{Curtis:1993py,Bloch:2001wz,Bloch:2002eq}. It is known that, the systematic treatment of overlapping divergences   hinges on a subtle interplay between the  multiplicative renormalization constant, $Z_c^{-1}$, and crucial 
contributions originated from the transverse part of the quark-gluon vertex. Since $\Gamma_\mu^{\Tr}$ is completely undetermined in our treatment, this delicate cancellation is already compromised. In particular, it is known that if, in addition to setting $\Gamma^{\Tr}_\mu = 0$, one uses the simplifying assumption that $Z_c^{-1} = 1$ in Eq.~\eqref{rsenergy1}, the resulting  anomalous  dimension of  the  quark mass  is incorrect.

A workaround for this problem was devised in~\cite{Aguilar:2010cn}, on the lines of an earlier proposal put forth in~\cite{Fischer:2003rp}. Namely, it consists in carrying out the substitution
\be
Z_c^{-1}{\mathcal K}_{\rm{\s {A,B}}}(p,k) \to
 {\mathcal K}_{\rm{\s{A,B}}}(p,k){\mathcal C}(q)\,,
\label{replace}
 \ee
 where  the function ${\mathcal C}(q)$  must be constructed in such a way that the product
\be\label{RGI}
{\mathcal R}(q)= \alpha_s(\mu)\Delta(q,\mu)F(q,\mu){\mathcal C}(q,\mu)\,,
\ee
is a renormalization group invariant (RGI), \mbox{($\mu$-independent)} combination, at least at one loop.

The requirement that ${\mathcal R}(q)$ be RGI completely fixes the ultraviolet behavior of ${\mathcal C}(q)$, namely
\bea
{\mathcal C}_{\rm{\s{UV}}}(q) = \left[1+\frac{9C_{\rm{\s A}} \alpha_s}{48\pi}\ln\left(\frac{q^2}{\mu^2}\right)\right]^{-1}\,,
\label{FUV}
\eea  
for large $q^2$. On the other hand, the infrared form of ${\mathcal C}(q)$ remains unspecified. 

The simplest function which displays the UV tail prescribed by Eq.~\eqref{FUV} is the ghost dressing function, $F(q)$, which is well understood in the infrared from lattice and continuum studies. As such, $F(q)$ is a natural candidate to play the role of $\mathcal{C}(q)$. However, since the criterion above does not determine ${ \mathcal C}(q)$ univocally, it is important to consider alternative infrared completions to Eq.~\eqref{FUV}, differing both quantitatively and qualitatively from the ghost dressing function.

\begin{figure*}[]
\begin{minipage}[b]{0.42\linewidth}
\centering
\includegraphics[scale=0.35]{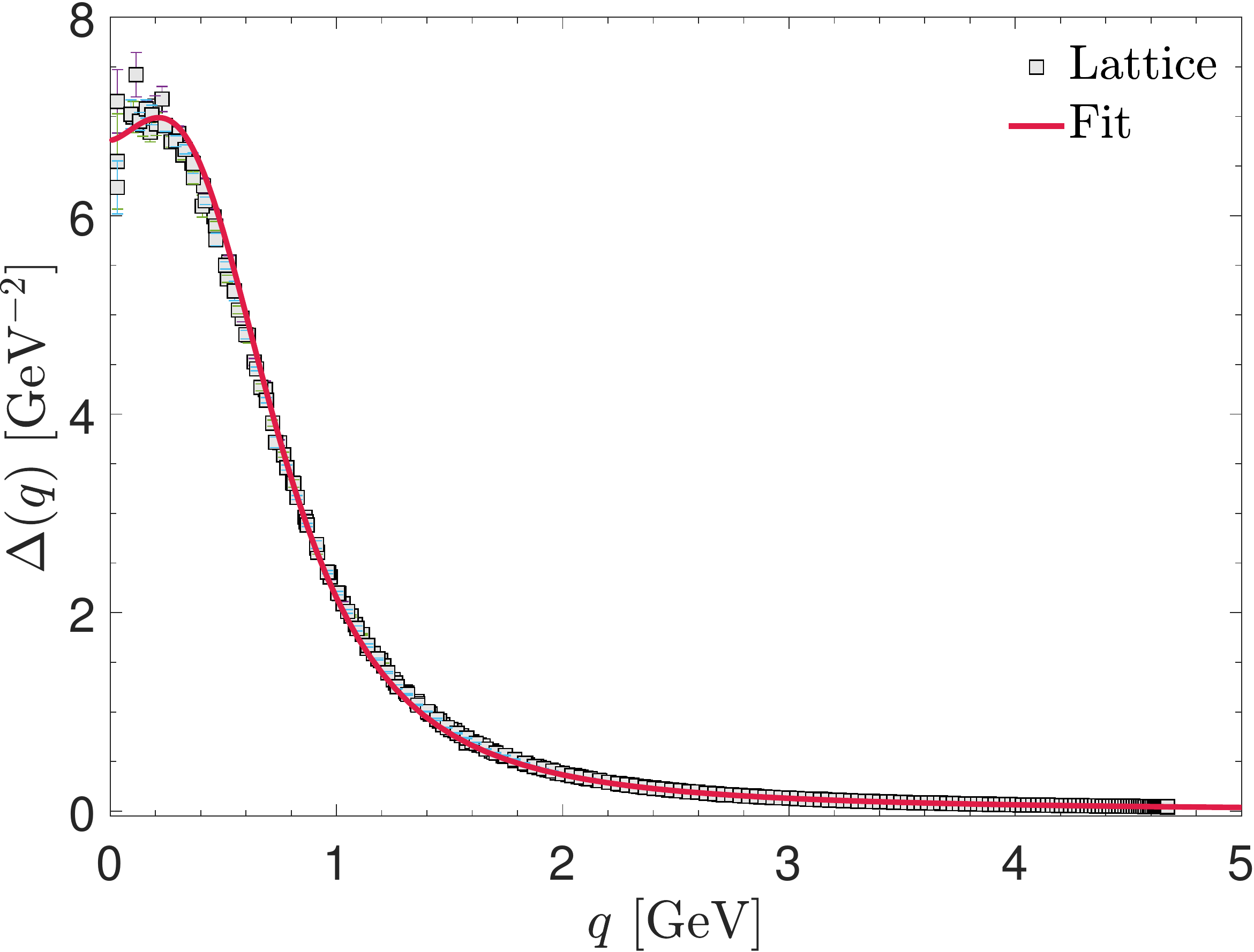}
\end{minipage}
\hspace{1.0cm}
\begin{minipage}[b]{0.42\linewidth}
\includegraphics[scale=0.35]{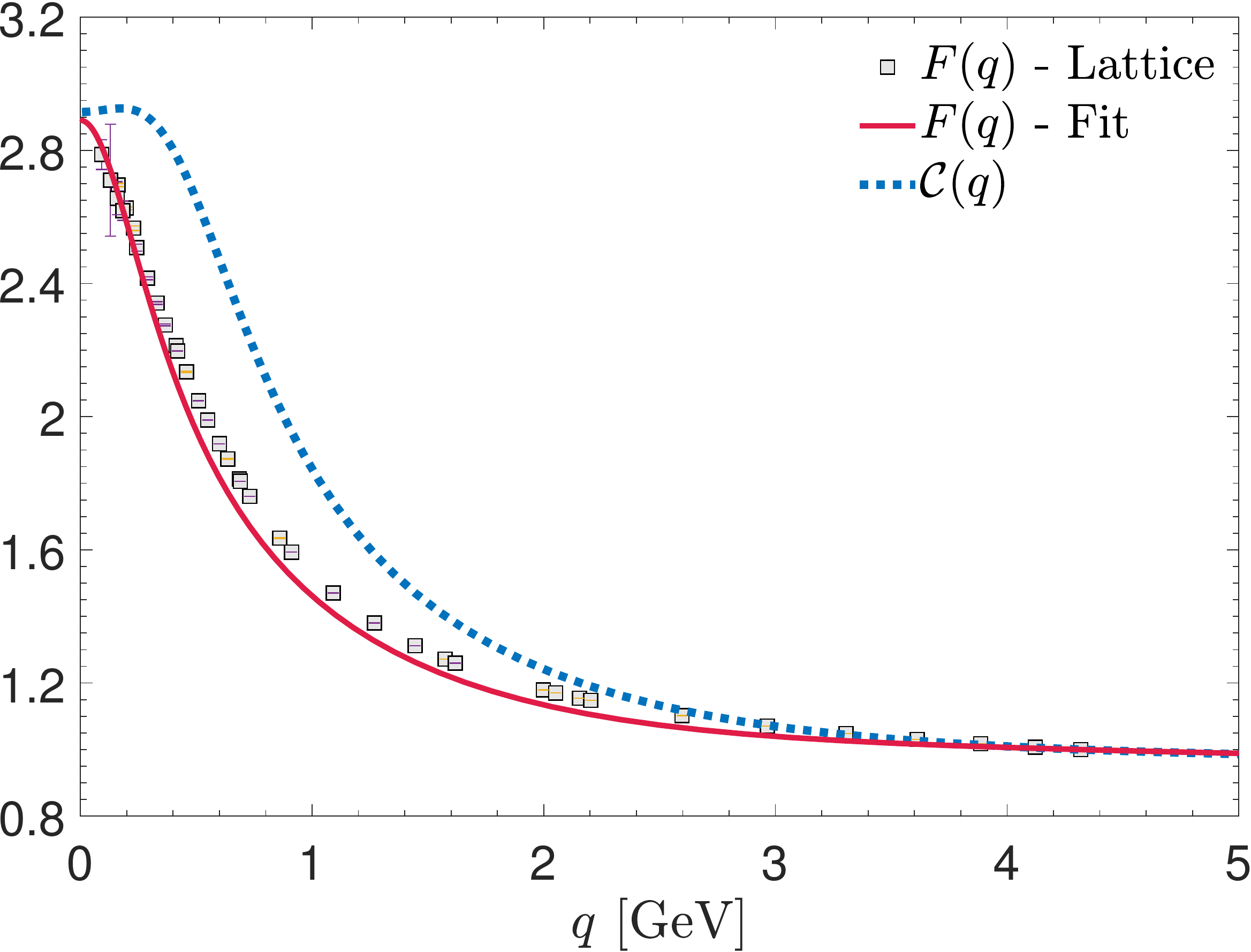}
\end{minipage}
\vspace{-0.3cm}
\caption{\label{func_ren} Lattice data for the gluon propagator, $\Delta(q)$, (left panel) and the ghost dressing function, $F(q)$, (right panel), from Ref.~\cite{Bogolubsky:2007ud}, and their respective fits (red continuous curves). In the right panel we also show the functions ${\mathcal C}(q)$ (blue dotted) given by the inverse of Eqs.~\eqref{1G}. The renormalization point in all cases is \mbox{$\mu = 4.3$\,GeV}.}
\end{figure*}

Another function that displays the one loop behavior required by Eq.~\eqref{FUV} is the inverse of the ``ghost-gluon mixing self-energy'', $[1 + G(q)]^{-1}$, which plays a key role in the pinch-technique~\cite{Binosi:2009qm,Aguilar:2008xm,Binosi:2014aea}, and equals the Kugo-Ojima function in Landau gauge~\cite{Kugo:1979gm,Grassi:2004yq,Fischer:2006ub,Aguilar:2009pp}. The $[1 + G(q)]^{-1}$ has the ultraviolet tail given by Eq.~\eqref{FUV}, while for low and intermediate momenta it can be determined through SDEs~\cite{Aguilar:2009nf}. In fact, thanks to the identity $[1 + G(0)]^{-1} = F(0)$, valid in Landau gauge~\cite{Aguilar:2009nf}, both functions
coincide at zero momentum, differing quantitatively only for intermediate momenta (see Fig.~\ref{func_ren}).

The SDE solutions for $1 + G(q)$ can be accurately fitted by the form~\cite{Aguilar:2018epe}
\be
 1+G(q) =  1+\frac{9C_{\rm{\s A}}\alpha_s}{48\pi}I(q)\ln\left(\frac{q^2+\rho_3 m^2(q)}{\mu^2}\right)  \,, 
 \label{1G}
 \ee 
where
 \bea
m^2(q)=\frac{m^4}{q^2+ \rho_2 m^2}\,; \;\;
I(q)= 1+ D \exp{\left(-\frac{\rho_4 q^2}{\mu^2}\right)} \nonumber \,,
\eea
and the fitting parameters are given by
\mbox{$m^2=0.55\,\mbox{GeV}^2$},  
\mbox{$\rho_2=0.60$}, 
\mbox{$\rho_3=0.50$},
\mbox{$\rho_4=2.08$},
\mbox{$\alpha_s=0.22$},
\mbox{$D=3.5$}, and  
\mbox{$\mu=4.3$\,GeV}.

For the purposes of this presentation, we will restrict ourselves to the analysis of the case where 
\mbox{${\mathcal C}(q)= [1+G(q)]^{-1}$}. Other functional forms for ${\mathcal C}(q)$ were  explored in more details in Ref~\cite{Aguilar:2018epe}.

Finally, after performing the substitution prescribed in Eq.~\eqref{replace} into Eq.~\eqref{rsenergy1}, we obtain the final versions of the integral equations for $A(p)$ and $B(p)$,
\bea\label{gAB1}
p^2A(p)  &=& p^2 + 4\pi C_{\rm{\s F}} \int_{k} {\mathcal K}_{\rm{\s A}}(k,p){\mathcal R}(q)\, , \nonumber\\
 B(p)  &=& 4\pi C_{\rm{\s F}}\int_{k} {\mathcal K}_{\rm{\s B}}(k,p){\mathcal R}(q)\,,
\eea
where $\mathcal{R}(q)$ is the RGI product defined in the Eq.~\eqref{RGI}.

Now, let us focus on the form factors of the scattering kernel $H(q,k,-p)$. 
The starting point in deriving the dynamical equations governing the behavior
of the  $X_i$ is the diagrammatic representation 
of $H^{[1]}(q,k,-p)$ at the one-loop dressed approximation, shown 
in  the bottom part of  Fig.~\ref{fig:system}, and written as 
\begin{align}
\!\!H^{[1]} =& 1 - \frac{i}{2}C_{\rm{\s A}}g^2\!\!\int_{l}\!\! \Delta^{\mu\nu}(l-k)D(l-p)G_{\mu}S(l)\Gamma_\nu\,,
\label{eqH_no_approx}
\end{align}
where $C_{\rm{\s A}}$ is the Casimir eigenvalue for the adjoint representation, and $G_{\mu}(p-l)$ is the ghost-gluon vertex.

Nonetheless, to proceed further with the derivation, we still need to truncate the vertices $G_\nu$ and $\Gamma_\mu$ appearing in the above equation. For the ghost-gluon vertex, we use simply its tree level form \mbox{$G^{(0)}_\nu= (p - l)_\mu$}, while for  the
quark-gluon vertex we will retain only the abelianized form factor $L_1^{BC}(l-k,k,-l)$, given in Eq.~\eqref{BC}. With the above simplifications, one has~\cite{Aguilar:2016lbe}
\bea
H^{[1]} =& 1- \frac{i}{2}C_{\rm{\s A}}g^2 \!\!\int_{l}\!\! \Delta^{\mu\nu}(l-k)(p-l)_{\mu}
D(l-p)S(l) L_1^{\rm \s{BC}}\gamma_{\nu}\,. \nonumber 
\label{eqH}
\eea

To obtain the equations for the individual $X_i$, one then contracts the 
above equation with the projectors defined in Eq.~(3.9) of Ref.~\cite{Aguilar:2016lbe}, yielding
\begin{align}
\label{generalx}
X_0 =& 1+ \lambda \!\int_l\mathcal{K}_{H}A(l){\mathcal G}(k,q,l)\,\\
X_1 =& \lambda \int_l\frac{\mathcal{K}_{H}B(l)}{q^2 h(p,k)} \left[k^2{\mathcal G}(p,q,l)-(p\cdot k){\mathcal G}(k,q,l)\right]\,,\nonumber\\
X_2 =& \lambda \int_l\frac{\mathcal{K}_{H}B(l)}{q^2 h(p,k)} \left[p^2{\mathcal G}(k,q,l)-(p\cdot k){\mathcal G}(p,q,l)\right] \,,\nonumber\\
X_3 =& -i \lambda \int_l\frac{\mathcal{K}_{H}A(l)}{q^2 h(p,k)} \nonumber \\
&\times\left[ k^2{\mathcal G}(p,q,l) - (p\cdot k){\mathcal G}(k,q,l)-{\mathcal T}(p,k,l)  \right]\,,\nonumber
\end{align}
where we define \mbox{$\lambda :=i \pi C_{\rm{\s A}}\alpha_s$} and  the kernel
\begin{align}
\mathcal{K}_{H}&=\frac{F(l-p)\Delta(l-k)[A(l)+A(k)]}{(l-p)^2[A^2(l)l^2-B^2(l)]}\,,
\label{kernelsH}
\end{align}
in addition, we have introduced the shorthand notation
\begin{align}
{\mathcal G}(r,q,l) =& (r\cdot q)-\frac{[r\cdot(l -k)][q\cdot(l-k)]}{(l-k)^2} \,,  \\
{\mathcal T}(p,k,l) =& (k\cdot q)[(p\cdot l) - (p\cdot k)] - (p\cdot q)[(k\cdot l)-k^2] \,.\nonumber
\label{fgg}
\end{align}

It is important to stress that the set of Eqs~\eqref{expLi} and~\eqref{generalx} for  $L_i$ and $X_i$, respectively, are written in Minkowski space, but may be converted to the Euclidean space using the rules stated in subsection {\rm III A} of~\cite{Aguilar:2016lbe}.

\begin{figure*}[]
\begin{minipage}[b]{0.42\linewidth}
\centering
\includegraphics[scale=0.35]{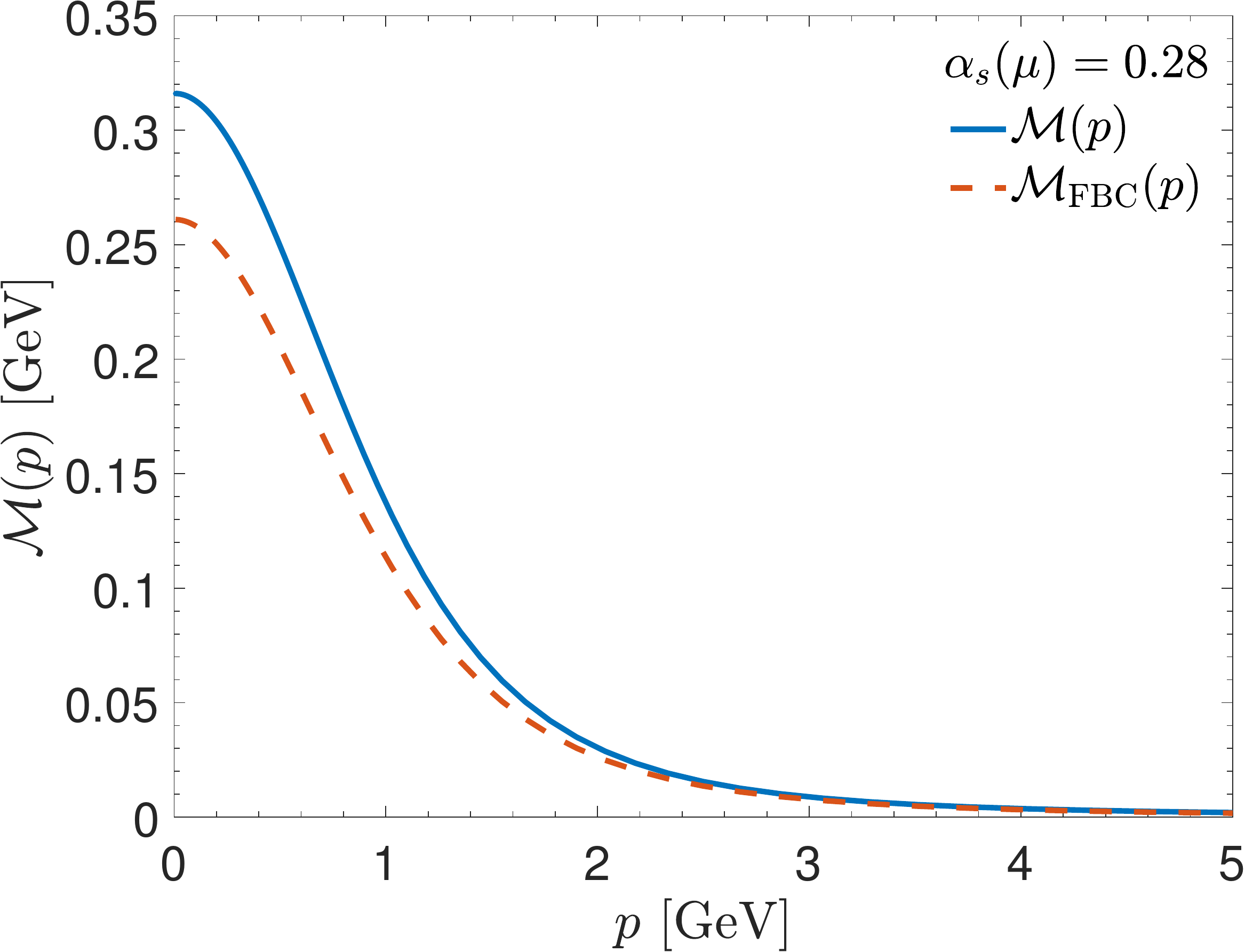}
\end{minipage}
\hspace{1.0cm}
\begin{minipage}[b]{0.42\linewidth}
\includegraphics[scale=0.35]{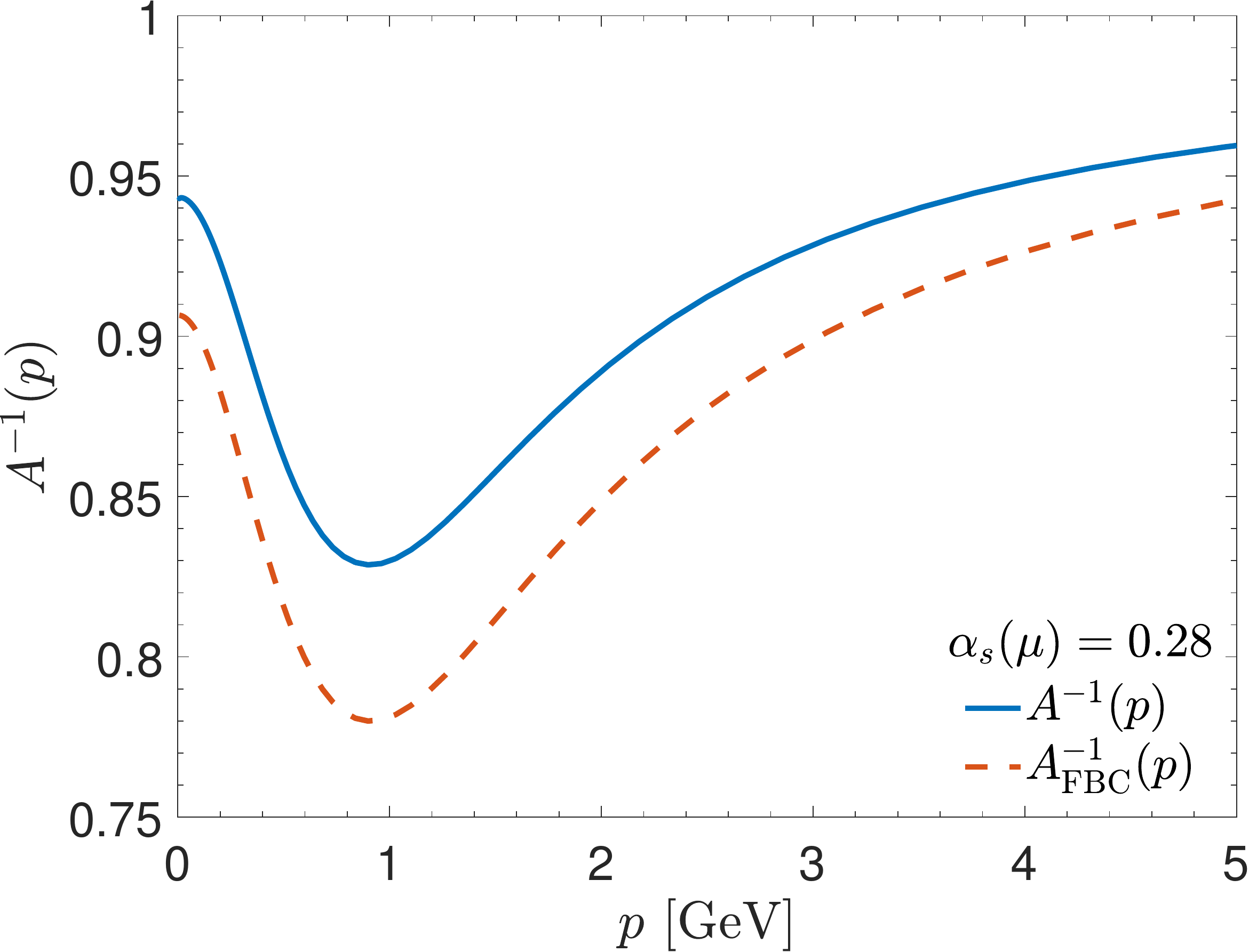}
\end{minipage}
\vspace{-0.3cm}
\caption{\label{mass_1G} Left panel: Comparison of ${\mathcal M}(p)$ obtained when the quark-gluon vertex employed into the quark gap equation is either $\Gamma^{\STI}_{\mu}$ (blue continuous) or $\Gamma^{\FBC}_{\mu}$ (orange dashed). Right panel: Same comparison, but for the $A^{-1}(p)$.}
\end{figure*}

\section{\label{sec:Num} Numerical analysis}

The truncated SDEs given by Eqs.~\eqref{gAB1}, \eqref{generalx} and the STI solution given by Eq.~\eqref{expLi}, comprise a coupled system of nonlinear equations, which is not closed only due to the need to specify  $\Delta(q)$, $F(q)$, and $\mathcal{C}(q)$. In principle, one could envisage further coupling the above six equations to the SDEs governing the behavior of $\Delta(q)$ and $F(q)$. However, the complexity of that approach would be too high. Instead, as done 
in a series of previous works~\cite{Aguilar:2010cn,Aguilar:2016lbe}, we close the system of equations considering 
 $\Delta(q)$, $F(q)$ and  $\mathcal{C}(q)$ as external ingredients. For that, we will employ for
  them suitable fits to lattice results obtained in the Ref.~\cite{Bogolubsky:2007ud}. The corresponding curves for $\Delta(q)$, $F(q)$, and $\mathcal{C}(q)$, renormalized at \mbox{$\mu = 4.3$\,GeV},  are shown in the Fig.~\ref{func_ren}.

With the above external ingredients, we are in position to solve numerically
the coupled system of  six integral equations for $A(p)$, $B(p)$, and the
four $X_i$ defined in the Eqs.~\eqref{gAB1} and \eqref{generalx}.

Given that the main feature of our truncation scheme is the
presence of the nontrivial contribution of $H$, expressed by the 
set of equations for the $X_i$, it will be interesting to assess the impact that $H$ has on the dynamical mass generation phenomenon. In Fig.~\ref{mass_1G}, we compare the result for the dynamical mass, ${\mathcal M}(p)$, and the quark wave function, $A^{-1}(p)$, obtained when we employ in the  gap equation either the full $\Gamma^{\STI}_{\mu}$ (blue continuous curves) or the ``minimally non-abelianized'' $\Gamma^{\FBC}_{\mu}$ (orange dashed ones). The
numerical solutions were obtained fixing $\alpha_s = 0.28$. For
a detailed analysis about the impact of $\alpha_s$ on the numerical solutions see Ref.~\cite{Aguilar:2018epe}. While, it is clear from Fig.~\ref{mass_1G} that the two solutions are qualitatively similar, the nontrivial contribution of $H$ produces a significant quantitative effect. In particular, the value of ${\mathcal M}(0)$ is about $21\,\%$ larger than ${\mathcal M}_{\FBC}(0)$~\cite{Aguilar:2018epe}. Therefore, the inclusion of $H$ into $\Gamma_{\mu}$ seems to be crucial to generate phenomenological compatible quark masses of the order~\mbox{$300$ MeV}.

\begin{figure*}[]
\begin{minipage}[b]{0.40\linewidth}
\centering
\includegraphics[scale=0.37]{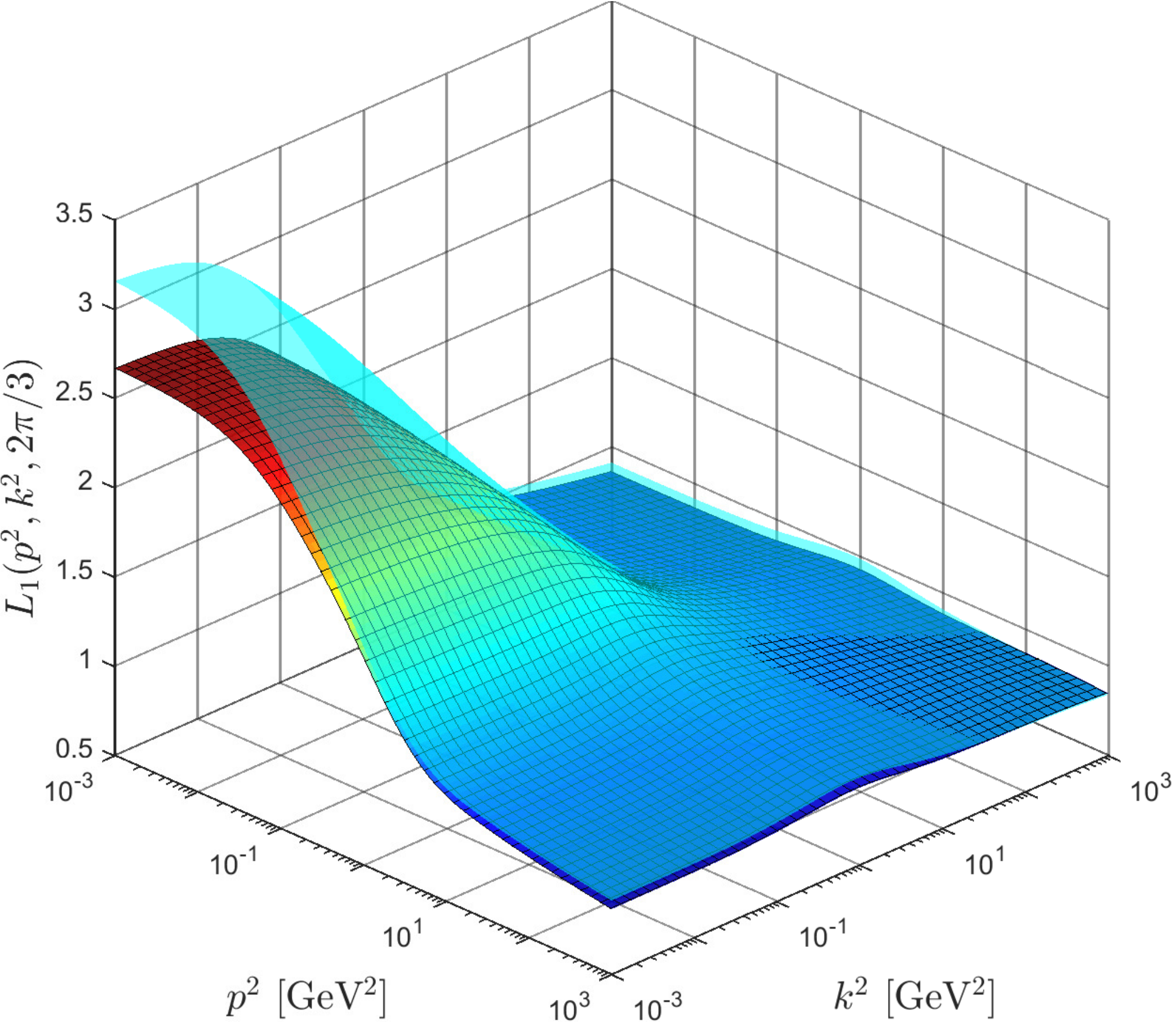}
\end{minipage}
\hspace{1.0cm}
\begin{minipage}[b]{0.45\linewidth}
\includegraphics[scale=0.37]{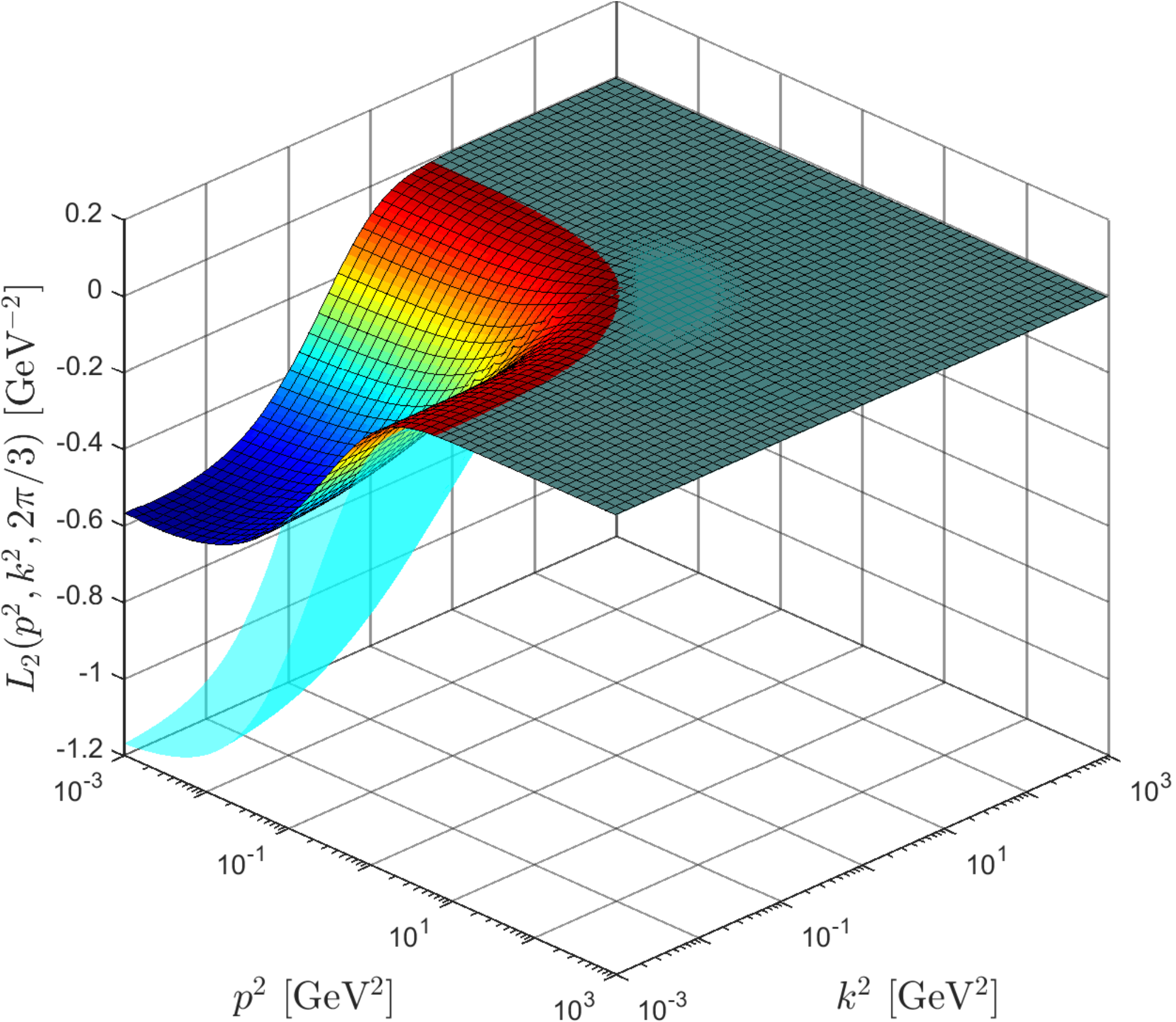}
\end{minipage}
\begin{minipage}[b]{0.40\linewidth}
\centering
\includegraphics[scale=0.37]{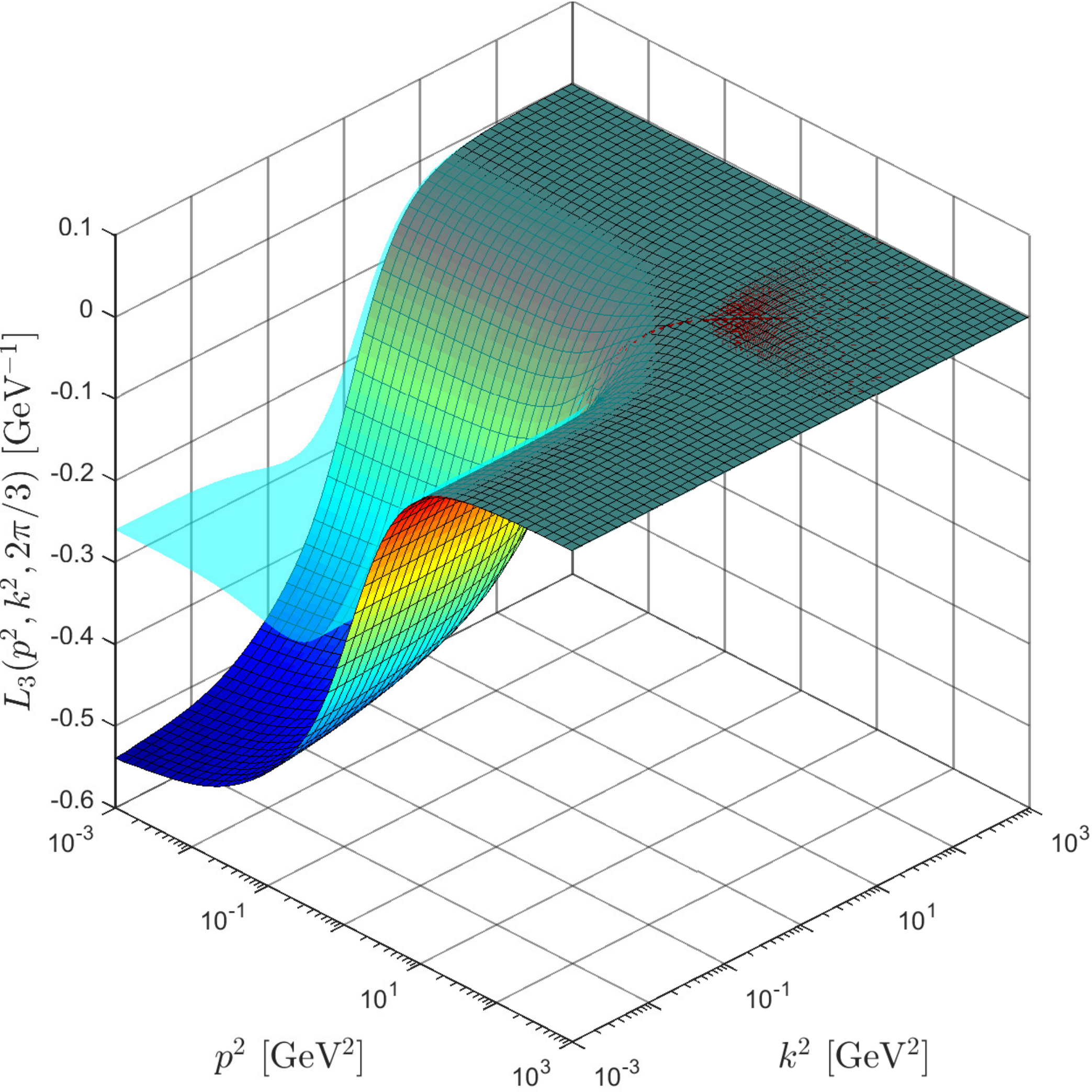}
\end{minipage}
\hspace{1.0cm}
\begin{minipage}[b]{0.45\linewidth}
\includegraphics[scale=0.36]{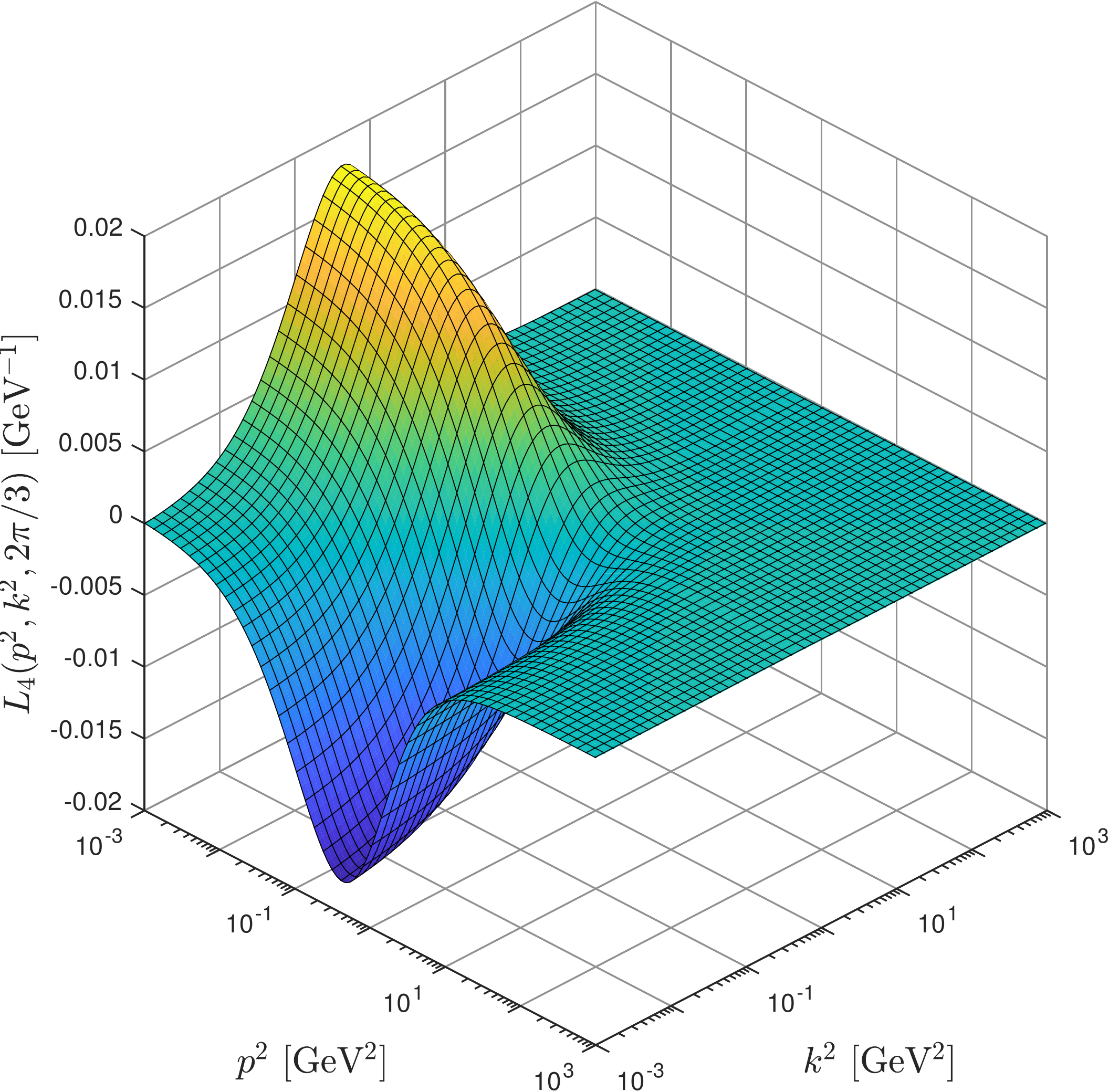}
\end{minipage}
\vspace{-0.3cm}
\caption{\label{fig:LSDE}  The  quark-gluon form factors $L_i$
obtained by substituting into  Eq.~\eqref{expLi} the solutions of the coupled 
system formed by Eqs.~\eqref{gAB1} and~\eqref{generalx}. The results
represent the case where $\alpha_s=0.28$  and 
$\theta=2\pi/3$.}
\end{figure*}
\begin{figure*}[]
\begin{minipage}[b]{0.42\linewidth}
\centering
\includegraphics[scale=0.35]{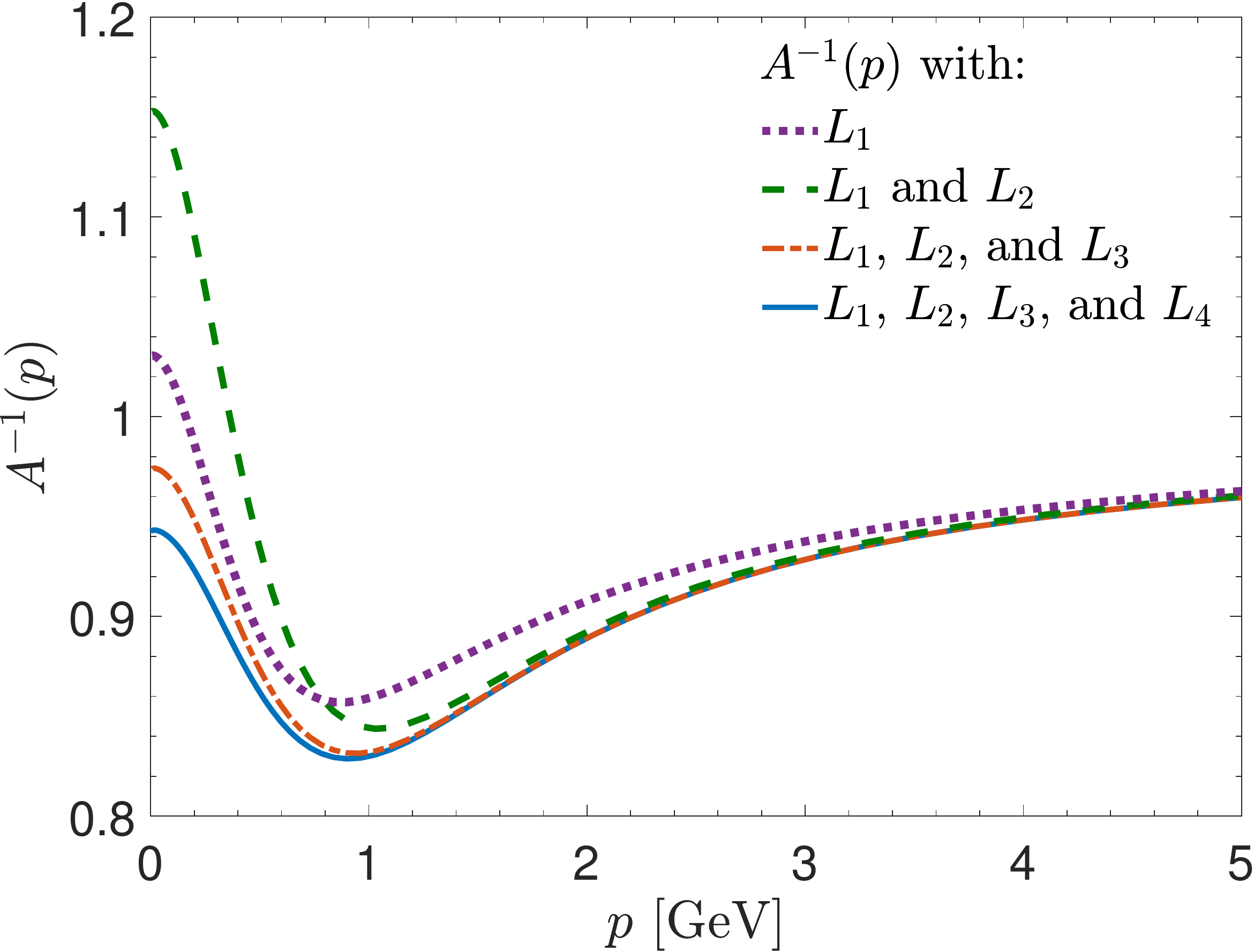}
\end{minipage}
\hspace{1.0cm}
\begin{minipage}[b]{0.42\linewidth}
\includegraphics[scale=0.35]{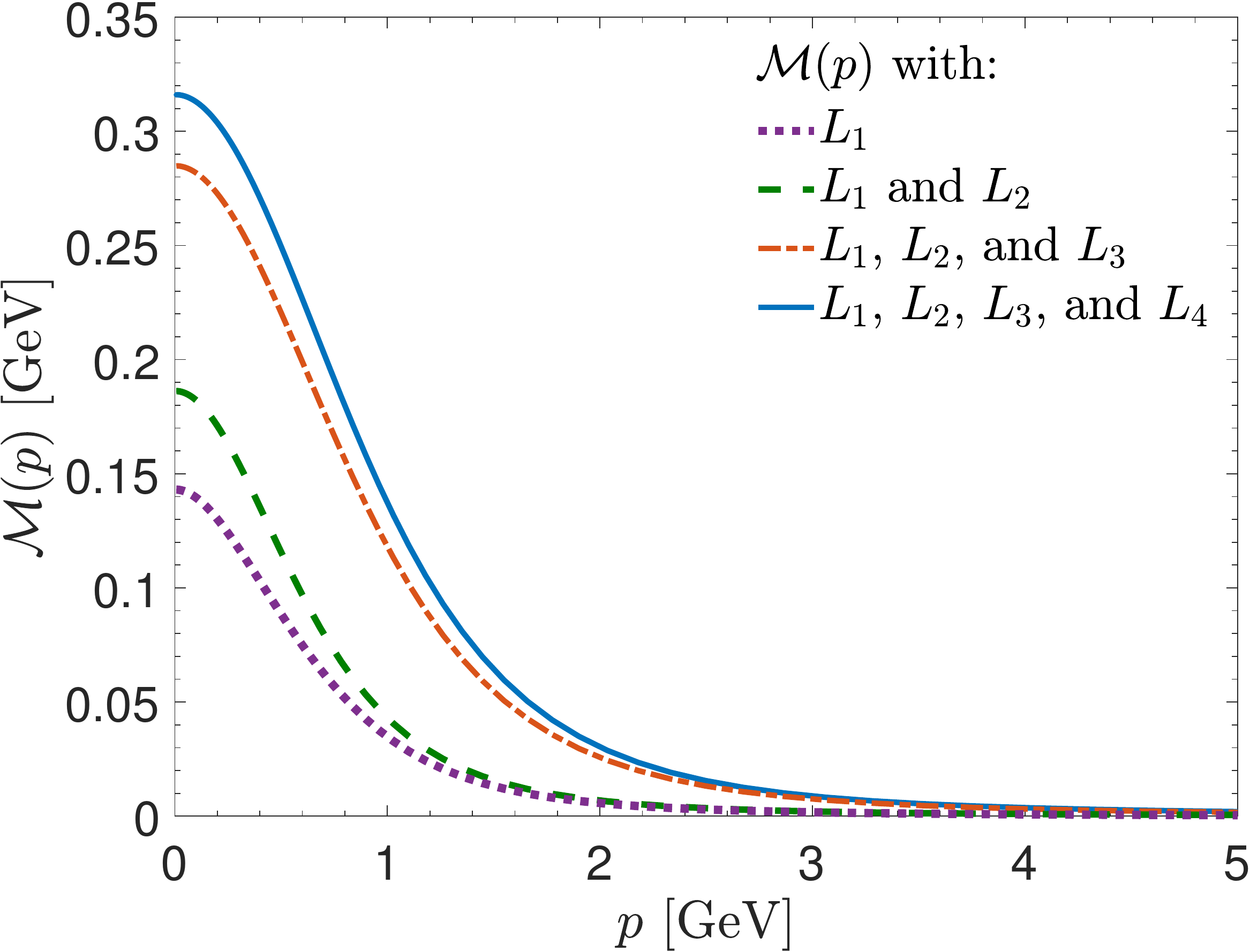}
\end{minipage}
\caption{\label{steps} Quark wave function, $A^{-1}(p)$, (left panel) and dynamical quark mass ${\mathcal M}(p)$ (right panel), obtained as the individual form factors are turned on sequentially. The legend for each curve says which $L_i$ are considered non zero during the computation.}
\end{figure*}

With the results for  $X_i$ at hands, we determine the corresponding form factors $L_i$, using the Euclidean version of Eq.~\eqref{expLi}. In Fig.~\ref{fig:LSDE}, we present a representative  set of results for 
the form factors $L_i$, where 
$\alpha_s=0.28$  and $\theta=2\pi/3$. Notice that $L_i$, represented by the colorful surfaces, 
display sizable deviations from the $L_i^{\FBC}$ represented by the cyan surface, given by Eq.~\eqref{BC}~\cite{Aguilar:2018epe}. It 
is also interesting to observe that $L_4$, contrary to $L_4^{\FBC}=0$, is a non-vanishing quantity, although its size is considerably suppressed  for all momenta.

Next, we turn our attention to the numerical impact of each individual $L_i$ on the results for ${\mathcal M}(p)$ and $A^{-1}(p)$. To this end, we solved the system of SDEs turning on gradually the $L_i$,  appearing in the
kernels of the gap equation~\eqref{gAB1}, starting with $L_1$ only.

The results of this analysis are shown in Fig.~\ref{steps}. While it is clear that indeed the form factor $L_1$ provides the largest contribution to the dynamical mass, it is interesting to notice that all of the $L_i$ contribute significantly to the strength of the kernel in the gap equation. More
specifically, $L_2$ furnishes $13\%$ of the ${\mathcal M}(0)$ value, while  
$L_3$ contributes another $23\%$. Quite surprisingly, the form factor $L_4$, often neglected in similar studies~\cite{Aguilar:2010cn,Fischer:2003rp,Roberts:1994dr}, provides about $10\,\%$ of the final ${\mathcal M}(0)$, despite being very small for all kinematic configurations (see Fig.~\ref{fig:LSDE}).

It is also interesting to mention that the quark running mass  
${\mathcal M}(p)$,
represented by the blue continuous curve in the Fig.~\ref{mass_1G},
may be accurately fitted by 
\bea
{\mathcal M}(p) = \frac{{\mathcal M}_1^3}{{\mathcal M}_2^2 + p^2\left[\ln(p^2+{\mathcal M}_3^2)/{\Lambda^2}\right]^{1-\gamma_f}} \,,
\label{fit2}
\eea
where the adjustable parameters are ~\mbox{${\mathcal M}_1 =758$ MeV}, \mbox{${\mathcal M}_2 =1.18$ GeV}, \mbox{${\mathcal M}_3 =426$ MeV}, and  \mbox{$\Lambda=270\,\mbox{MeV}$}.  Notice that the  presence of the ${\mathcal M}_2$ in the denominator enforces the saturation of ${\mathcal M}(p)$
at the origin, while the ${\mathcal M}_3$  in the argument of the logarithm 
 improves the convergence of the fitting procedure. 

Finally, we have used the pion decay constant, $f_{\pi}$, 
to assess  the impact that the
inclusion of $H$ in the construction of the $\Gamma^{\STI}_{\mu}$ might have on physical quantities.  Using an improved version of the Pagels-Stokar-Cornwall formula~\cite{Roberts:1994hh}, we obtain \mbox{$f_{\pi}=87$ MeV} when we employ the solutions represented by the orange dashed curve of Fig.~\ref{mass_1G}, while for 
$\Gamma^{\STI}_{\mu}$ (blue continuous curve of Fig.~\ref{mass_1G})  we obtain 
\mbox{$f_{\pi}=97$ MeV}. Therefore the final impact of $H$ is to increase approximately by $10\%$ the value of  \mbox{$f_{\pi}$}. We remind that the  above quoted values for \mbox{$f_{\pi}$} should be compared to the experimental value \mbox{$f_{\pi}=93$ MeV}~\cite{Patrignani:2016xqp}.

\vspace{-0.2cm}

\section{\label{sec:Conclusions} Conclusions}

\vspace{-0.2cm}
We have carried out a detailed study of the impact of the quark-ghost scattering kernel on the dynamical quark mass generation through a coupled system of equations composed by the quark propagator $S(p)$ and the one-loop dressed truncation for $H$. In the truncation 
scheme adopted, we have neglected the transverse part of the quark-gluon vertex, $\Gamma_\mu$, which cannot be determined from the STI that this vertex  satisfies.

Our results demonstrate that the inclusion of a nontrivial contribution of $H$ in the construction of $\Gamma^{\STI}_\mu$ has a substantial quantitative effect on the infrared behavior of the  quark propagator. Particularly important is the effect on the dynamical mass, which increased by about $20\,\%$ in comparison to the result obtained with the ``minimally non-abelianized'' vertex, $\Gamma^{\FBC}_\mu$. 

A surprising result of our analysis is that the form factor $L_4$ contributed about $10\,\%$ to the total dynamical mass, in spite of its rather suppressed
structure compared to $L_1$, $L_2$ and $L_3$~\cite{Aguilar:2016lbe}. This can be explained by the fact that this form factor peaks in the region of momenta around $1\text{ GeV}$ (see Fig.~\ref{fig:LSDE}), where the support of the gap equation kernel seems to be most critical~\cite{Fischer:2003rp,Aguilar:2010cn,Roberts:1994dr,Maris:1999nt}.

Lastly, the difficulties in enforcing multiplicative renormalizability at the level of the gap equation, and the subsequent restoration of the correct anomalous dimension for the quark  dynamical mass was circumvented by the introduction, by hand, of a function ${\mathcal C}(q)$ to correct the UV behavior of the gap equation kernel. However, this procedure is ambiguous in what regards the IR completion of ${\mathcal C}(q)$. By solving the system of equations with different forms for ${\mathcal C}(q)$ (see Ref.~\cite{Aguilar:2018epe} for more details), we found more evidence that the support of the kernel of the gap equation in the region of momenta around $1 \text{ GeV}$ is crucial for the generation of phenomenologically compatible quark masses.

Nevertheless, a consistent determination of the transverse part of the quark-gluon vertex, is mandatory in order to  better understand the renormalizability of the gap equation.

\vspace{-0.4cm}

\acknowledgments 
\vspace{-0.5cm}

The authors thank the organizers of the {\rm XIV} International Workshop on Hadron Physics for their hospitality. The work of  A.~C.~A and M.~N.~F. are supported by CNPq  under the grants 305815/2015 and 142226/2016. respectively. A.~C.~A also acknowledges the financial support from FAPESP through the projects  2017/07595-0 and  2017/05685-2. This research was performed using the Feynman Cluster of the
John David Rogers Computation Center (CCJDR) in the Institute of Physics ``Gleb
Wataghin", University of Campinas.

\vspace{-0.7cm}


%

\end{document}